\documentclass[pra,aps,twocolumn,superscriptaddress,showpacs]{revtex4}
\usepackage{graphicx}
\usepackage{amsmath}

\newcommand{\ket}[1]{|#1\rangle}
\newcommand{\bra}[1]{\langle#1|}

\newcommand{\eq}{\begin{equation}}
\newcommand{\fine}{\end{equation}}

\begin{document}

\title{

\bf \LARGE  Rigorous Quantum Limits on Monitoring Free Masses and Harmonic Oscillators. }
\vspace{.2cm}
\date{5 February, 2017}

\author{S. M. Roy}
\email{smroy@hbcse.tifr.res.in} \affiliation{HBCSE,Tata Institute of Fundamental Research, Mumbai}

\begin{abstract}
There are heuristic arguments proposing 
that the accuracy of monitoring position of a free mass $m$ is limited by the standard quantum limit 
(SQL) (\cite{Braginsky},\cite{Caves1980}):$\sigma^2 (X(t)) \geq  \sigma^2 (X(0)) +(t^2/m^2) \sigma^2 (P(0))\geq \hbar t/m$, where $\sigma^2 (X(t))$ 
and $\sigma^2 (P(t))$ denote variances of the Heisenberg representation position and momentum operators. Yuen \cite{Yuen}
 discovered that there are contractive states for which this result is incorrect. Here I prove universally valid rigorous quantum limits (RQL)
viz. rigorous upper and lower bounds on $\sigma^2 (X(t))$ in terms of $\sigma^2 (X(0))$ and $\sigma^2 (P(0))$  given by Eqn. (\ref{RQL}) for a free mass, 
and by Eqn. (\ref{RQLOSCXP}) for an oscillator. 
I also obtain the `maximally contractive' and `maximally expanding' states which saturate the RQL, and use the contractive states to set up  
an Ozawa-type \cite{Ozawa1988} measurement theory with accuracies respecting the RQL but beating the standard quantum limit.
The Contractive states for oscillators  improve on the Schr\"odinger coherent states of constant variance and may be useful 
for gravitational wave detection and optical communication.

\end{abstract}
 
\pacs{03.65.-W , 03.65.Ta ,04.80.Nn}

\maketitle

{\bf Introduction}. A quantum system is prepared, for example by a measurement, in an initial state . 
Subsequent monitoring or measurements of an observable $A$ may be useful to detect any external 
disturbances additional to the intrinsic change in the uncertainty of the observable due to the system evolving by its 
own Hamiltonian.Much before the actual discovery of gravitational waves \cite{Abbott} it was realised that accurate monitoring of 
position of an oscillator and of a free mass, including quantum effects are important for 
gravitational wave interferometers \cite{Thorne}.

For an arbitrary initial state of a free mass or an oscillator , I shall obtain rigorous quantum 
limits (RQL) on the intrinsic uncertainty after time $t$.

For any observable with Schr\"odinger operator $A$ (e.g. position $A=X$ or momentum $A=P$), and any Hamiltonian $H$, 
the Heisenberg operator $A(t)$ at time t and its variance $\sigma^2 (A(t)) $ are defined by,
\begin{eqnarray}
&& A(t)\equiv exp(iHt/\hbar)\> A \>exp (-iHt/\hbar),\> \\
&& \sigma^2 (A(t)) \equiv \bra{\psi(0)}(\Delta A(t))^2 \ket {\psi(0)},\\
&& \Delta A(t)\equiv A(t)-<A(t)>,\\
&& <A(t)>\equiv\bra{\psi(0)}A(t)\ket{\psi(0)}
\end{eqnarray}
where $\ket {\psi(0)}$ is the initial state.

{\bf Heuristic Standard Quantum Limit on Monitoring Position of a Free Mass .}
There are heuristic arguments proposing 
that the accuracy of position monitoring is limited by the standard quantum limit (SQL) (\cite{Braginsky},\cite{Caves1980})
on the variance of the position operator $X(t)$ :

\begin{eqnarray} \label{SQL}
 &\sigma^2 (X(t)) \geq  \sigma^2 (X(0)) +(t^2/m^2) \sigma^2 (P(0)) \\
 &\geq  2(t/m)\sigma(X(0))\sigma (P(0)) \geq \hbar t/m, 
\end{eqnarray}

 For the free mass , $H=P^2/(2m) $. the inequality (\ref{SQL}) is actually an equality for Gaussian states,

\begin{eqnarray}
 <p\ket {\psi (t)} = (\pi \alpha )^{-1/4} exp [-\frac {(p-\beta )^2 } {2\alpha } -it\frac{p^2 } {2m } ],\nonumber \\
 \sigma ^2 ( P (t)) = \frac{\alpha}{2}, \> \sigma^2 (X(t))=\hbar ^2 \frac{1+ (\alpha t/(m\hbar))^2 } {2\alpha }.
\end{eqnarray}
 

 One heuristic argument for the SQL (\cite{Braginsky},\cite{Caves1980}) Eq. (\ref{SQL}) starts from 
$H=P^2/(2m),\Delta X(t)= \Delta X(0) +(t/m)\Delta P(0) $, 
\begin{eqnarray}\label{Heisenberg}
 \sigma^2 (X(t))=\sigma^2 (X(0))+ (t^2/m^2) \sigma^2 (P(0)) + \nonumber \\
 (t/m)\bra{\psi(0)} \Delta X(0) \Delta P(0) +\Delta P(0) \Delta X(0) \ket{\psi(0)}.
\end{eqnarray}
One obtains the SQL if one assumes that the third term on the right-hand side is non-negative.

In a seminal paper, Yuen \cite{Yuen} noted that there are contractive states for which this assumption is incorrect.
In an interesting and correct argument for the SQL, valid in  certain measurement models, Caves \cite{Caves1985} noted that 
in some models, resolution of the meter $\geq \sigma (X(0)) $  may entail that the variance of the 
position measurement at time $t$ is $\geq \sigma^2 (X(0)) + \sigma^2 (X(t))$ which is 
$\geq \hbar t/m$ by the uncertainty principle. Yuen \cite{Yuen} and Ozawa \cite{Ozawa1988} (see also \cite{Arthurs-Kelly}),
point out the existence of other measurement models for which the imperfect resolution correction can be much 
smaller than   $\sigma^2 (X(0))$.  I address myself first to finding a rigorous version 
of the heuristic SQL Eq. ( \ref{SQL}) on $\sigma^2 (X(t))$ and optimum contractive states. 
I then briefly discuss how the Ozawa\cite{Ozawa1988} measurement model and the contractive states may be used for 
repeated measurements on oscillators or free masses over finite times , respecting of course the 
rigorous quantum limits (RQL) presented here, but  beating the SQL .

 {\bf Rigorous Quantum Limit on Monitoring Position of a Free Mass .}
We start from Eq. (\ref{Heisenberg}) and find exact limits on the third term on the right-hand side.
Using 
\begin{equation}
 [\Delta X(0), \Delta P(0)]=i\hbar , 
\end{equation}
we have,
\begin{eqnarray}
 &\bra{\psi(0)} \Delta X(0) \Delta P(0) +\Delta P(0) \Delta X(0) \ket{\psi(0)}+i\hbar \nonumber\\
 &=2 \bra{\psi(0)} \Delta X(0) \Delta P(0)  \ket{\psi(0)}.
\end{eqnarray}
Cauchy-Schwarz inequality on the right-hand side yields,
\begin{eqnarray}\label{Cauchy}
 &\big(\bra{\psi(0)} \Delta X(0) \Delta P(0) +\Delta P(0) \Delta X(0) \ket{\psi(0)}\big)^2\nonumber\\
 &\leq 4 \sigma^2(X(0)) \sigma^2(P(0)) -\hbar ^2  ,
 \end{eqnarray}
 which is a rearrangement of the Schr\"odinger-Robertson   uncertainty relation on the product of variances of $X$ and $P$ 
 (\cite{Kennard},\cite{Mukunda}).
 
 Substituting this into Eq. (\ref{Heisenberg}) I have the rigorous quantum limits (RQL),
\begin{eqnarray}\label{RQL}
&& \sigma^2(X(0)) +(t/m)^2 \sigma^2(P(0)) \nonumber \\
&&-(t/m) \sqrt{4 \sigma^2(X(0))\sigma^2(P(0)) -\hbar^2 } \nonumber \\ 
&&\leq \sigma^2(X(t)) \nonumber \\
&& \leq \sigma^2(X(0)) +(t/m)^2 \sigma^2(P(0))\nonumber \\
&&+(t/m) \sqrt{4 \sigma^2(X(0))\sigma^2(P(0)) -\hbar^2 }.
\end{eqnarray}
{\bf It must be stressed that the bounds are fundamental quantum limits valid for arbitrary states}.
The only states saturating the inequalities are those for which the Schwarz inequalities are equalities, 
i.e. $ \Delta P(0)  \ket{\psi(0)}$  is a complex constant times $\Delta X(0)  \ket{\psi(0)}$. Hence the RQL ,Eq. (\ref{RQL})
are equalities if and only if,
\begin{eqnarray}
&& \Delta P(0)  \ket{\psi(0)} =i\lambda \Delta X(0)  \ket{\psi(0)} ,\\
&& <X' \ket{\psi (0)}= \big ( \frac{Re \lambda } {\pi \hbar } \big )^{1/4} \nonumber \\
&&\times exp \big(\frac{i<P(0)> X'} {\hbar } -\frac{\lambda (X'-<X(0)>)^2 } {2\hbar }\big ),\label{optimal state}
\end{eqnarray}
with $ Re \lambda > 0$, 
\begin{eqnarray}
 |Im \lambda | =  \frac{1}{2 \sigma^2(X(0))} \sqrt{4 \sigma^2(X(0))\sigma^2(P(0)) -\hbar^2 } ,\nonumber\\
 \sigma^2(X(0)) =\hbar/(2  Re \lambda) ,\>\sigma^2(P(0)) =\hbar | \lambda |^2 /(2  Re \lambda) ,
\end{eqnarray}

and,
\begin{eqnarray}
 \bra{\psi(0)} \Delta X(0) \Delta P(0) +\Delta P(0) \Delta X(0) \ket{\psi(0)} \nonumber\\
= \mp  \sqrt{4 \sigma^2(X(0))\sigma^2(P(0)) -\hbar^2 }, \>if \> Im \lambda = \pm |Im \lambda |
\end{eqnarray}

The  positive and negative signs of $Im \lambda $ correspond respectively to saturation of 
the left-hand side and right-hand side of the inequality(\ref{RQL}). 
The right-hand side of inequality (\ref{RQL}) sets an upper limit on spreading 
of the position wave packet and the left-hand side to the amount of contraction possible. The states (\ref{optimal state}) with 
positive $Im \lambda $ derived without any reference to oscillators turn out to be essentially Yuen's contractive 
Twisted Coherent States (TCS) \cite{Yuen} of an associated fictitious oscillator. Thus,the above demonstration shows that 
for given $\sigma(X(0)),\sigma(P(0)) $, the (TCS) are the optimum contractive states.

It is useful to rewrite the left-hand side of the inequality (\ref{RQL}) in two alternative forms:
\begin{eqnarray}
&& \sigma^2(X(t)) \ge \big(\frac{\hbar}{2\sigma (P(0))}\big)^2 +\big(\frac{\sigma (P(0))}{m} \big)^2 (t-\frac{1}{2}t_M)^2 \label{RQL1}\\
&&=\frac{t}{m} \big(2\sigma(X(0))\sigma(P(0))-\sqrt{4 \sigma^2(X(0))\sigma^2(P(0)) -\hbar^2 } \big)\nonumber\\
&&+\big( t\frac{\sigma (P(0))}{m}-\sigma(X(0)\big)^2 , \label{RQL2}
\end{eqnarray}
where ,
\begin{equation} \label{tM1}
 t_M= \frac{m}{ \sigma^2(P(0)) } \sqrt{4 \sigma^2(X(0))\sigma^2(P(0)) -\hbar^2 } .
\end{equation}
Eq.(\ref{RQL1}) shows that the optimal state (\ref{optimal state}) with positive $Im \lambda $ remains contractive upto time $t_M/2$, 
and the variance $\sigma^2(X(t))$ is less than the initial variance $\sigma^2(X(0))$ for time $t <t_M$, i.e.
\begin{equation}
 \sigma^2(X(t)) \leq \sigma^2(X(0)) ,\>for \> t \leq t_M ,
\end{equation}\label{tM1Max}
 for the optimum contractive state.
Eq.(\ref{RQL2}) shows that 
for a given uncertainty product, by choosing $(t/m) \sigma ^2 (P(0))=\sigma(X(0))\sigma (P(0))$, 
$\sigma^2(X(t))$ can be made as small as $(t/m) \big(2 (\sigma(X(0))\sigma(P(0))- \sqrt{4 \sigma^2(X(0))\sigma^2(P(0)) -\hbar^2 }\big)  $ ;
 this is $\approx t\hbar ^2 /(4m \sigma(X(0))\sigma(P(0)) )$ for a large uncertainty product, and can be much smaller than the heuristic standard 
 quantum limit $\hbar t/m $ .


{\bf Rigorous Quantum Limits on Monitoring Position or Momentum of a Harmonic Oscillator . }
This problem is specially significant because Hamiltonians for all free Bosonic fields , including the electromagnetic field , 
are sums of Harmonic oscillator Hamiltonians. In particular, the limits I derive can be immediately translated 
into rigorous quantum limits (RQL) on time development of quadratures of the electromagnetic field.

The Hamiltonian $ H=P^2/(2m) + \frac{1}{2} m\omega^2 X^2$ can be rewritten as,
\begin{equation}
 H=\frac{1}{2}\hbar \omega (p^2+x^2)=\hbar \omega (a^\dagger a +1/2),
\end{equation}
where,
\begin{eqnarray}
 p=\frac{P}{\sqrt{m\hbar \omega }},\>x=\sqrt{\frac{m\omega}{\hbar} }X ,\nonumber\\
 a=\frac{x+ip }{\sqrt{2} },\> a^\dagger=\frac{x-ip }{\sqrt{2} }.
\end{eqnarray}
The Heisenberg equations of motion yield,
\begin{eqnarray}
 && \Delta x(t)=\cos (\omega t)\> \Delta x(0) +\sin (\omega t)\>\Delta p(0)\nonumber\\
 && \Delta p(t)=-\sin (\omega t)\> \Delta x(0) +\cos (\omega t)\>\Delta p(0).
\end{eqnarray}
Hence,
\begin{eqnarray}
 \sigma^2 (x(t))=\cos ^2(\omega t)\sigma^2 (x(0))+  \sin ^2(\omega t)\sigma^2 (p(0)) \nonumber \\
+ \frac{1}{2} \sin (2\omega t)\bra{\psi(0)} \Delta x(0) \Delta p(0) +\Delta p(0) \Delta x(0) \ket{\psi(0)},
\end{eqnarray}
\begin{eqnarray}
 \sigma^2 (p(t))=\sin ^2(\omega t)\sigma^2 (x(0))+  \cos ^2(\omega t)\sigma^2 (p(0)) \nonumber \\
- \frac{1}{2} \sin (2\omega t)\bra{\psi(0)} \Delta x(0) \Delta p(0) +\Delta p(0) \Delta x(0) \ket{\psi(0)}.
\end{eqnarray}
As before, using $[\Delta x(0) ,\Delta p(0)]=i $, and Schwarz inequality, we obtain,

 \begin{eqnarray}
 &\big(\bra{\psi(0)} \Delta x(0) \Delta p(0) +\Delta p(0) \Delta x(0) \ket{\psi(0)}\big)^2\nonumber\\
 &\leq 4 \sigma^2(x(0)) \sigma^2(p(0)) -1 .
\end{eqnarray}
 Hence, we have the RQL for the oscillator in terms of the dimensionless variables $x$ and $p$ which 
 can be the quadratures for a mode of frequency $\omega$ of the electromagnetic field,
\begin{eqnarray}\label{RQLOSC1}
&&\big(\cos ^2(\omega t)\sigma^2 (x(0))+ \sin ^2(\omega t)\sigma^2 (p(0))\big)\nonumber\\
&&-\frac{1}{2} |\sin (2\omega t)| \sqrt{4 \sigma^2(x(0)) \sigma^2(p(0)) -1 } \nonumber\\
&&\leq  \sigma^2 (x(t)) \nonumber\\
&&\leq \big(\cos ^2(\omega t)\sigma^2 (x(0)) + \sin ^2(\omega t)\sigma^2 (p(0))\big)\nonumber\\
&&+\frac{1}{2} |\sin (2\omega t)| \sqrt{4 \sigma^2(x(0)) \sigma^2(p(0)) -1 }
\end{eqnarray}
which corresponds to Eqn.(\ref{RQL} ) for a free mass. We also have  RQL for $\sigma^2 (p(t)) $ for the oscillator,

 \begin{eqnarray}\label{RQLOSC2}
 &&\big(\sin ^2(\omega t)\sigma^2 (x(0))+ \cos ^2(\omega t)\sigma^2 (p(0))\big)\nonumber\\
&&-\frac{1}{2} |\sin (2\omega t)| \sqrt{4 \sigma^2(x(0)) \sigma^2(p(0)) -1 } \nonumber\\
&&\leq  \sigma^2 (p(t)) \nonumber\\
&&\leq \big(\sin ^2(\omega t)\sigma^2 (x(0))+  \cos ^2(\omega t)\sigma^2 (p(0))\big)\nonumber\\
&&+\frac{1}{2} |\sin (2\omega t)| \sqrt{4 \sigma^2(x(0)) \sigma^2(p(0)) -1 }.
\end{eqnarray}
The extremal states saturating these RQL may be written in terms of the dimensionless variables $x,p$ for 
use in optical quadrature measurements,
\begin{eqnarray}\label{extremalOSC}
&& \big(\Delta p(0)-i\eta_{\pm} \Delta x(0) \big ) \ket{\psi (0)_{\pm} }=0 \\ 
&& <x' \ket{\psi (0)_{\pm}}= \big ( \frac{Re \>\eta _{\pm}} {\pi } \big )^{1/4} \nonumber \\
&& \times \exp \big(i<p(0)> x') -\frac{\eta _{\pm}(x'-<x(0)>)^2 } {2}\big ),\label{chi}
\end{eqnarray}
with 
\begin{equation}\label{etavalues}
 \eta _{\pm} = \frac{1}{2 \sigma^2(x(0))}[1 \pm i \sqrt{4 \sigma^2(x(0))\sigma^2(p(0)) -1 }] .
\end{equation}
The values $\eta=\eta _{\pm}$ yield the values $ \sigma^2 (x(t))_{\pm}$ and $\sigma^2 (p(t))_{\pm} $,

\begin{eqnarray}\label{RQLOSC4}
 \sigma^2 (x(t))_{\pm}-\cos ^2(\omega t)\sigma^2 (x(0)) -  \sin ^2(\omega t)\sigma^2 (p(0))\nonumber \\
=\mp \frac{1}{2} \sin (2\omega t) \sqrt{4 \sigma^2(x(0)) \sigma^2(p(0)) -1 },
\end{eqnarray}
and 
 \begin{eqnarray}\label{RQLOSC5}
 \sigma^2 (p(t))_{\pm}-\sin ^2(\omega t)\sigma^2 (x(0))-  \cos ^2(\omega t)\sigma^2 (p(0)) \nonumber \\
=\pm \frac{1}{2} \sin (2\omega t) \sqrt{4 \sigma^2(x(0)) \sigma^2(p(0)) -1 }.
\end{eqnarray}
We deduce ,for example, that for the initial state $\ket{\psi (0)_+}$, 
\begin{equation}
\sigma^2 (x(t))_+\leq \sigma^2 (x(0)),if \>0\leq \omega t \leq \omega t_M ', 
\end{equation}\label{tM2Max}
where,
\begin{equation} \label{tM2}
\omega t_M ' \equiv \tan ^{-1} [\frac {\sqrt{4 \sigma^2(x(0)) \sigma^2(p(0)) -1 } }{\sigma^2 (p(0))- \sigma^2 (x(0))}] <\pi \>,
\end{equation}
which corresponds to Eq. (\ref{tM1}) in the free mass case.

Upto time $t_M'$ , the contractive states for the oscillator thus improve on the Schr\"odinger coherent states which have constant 
$ \sigma^2 (x(t))$. Analogous results are easily obtained for $ (\sigma^2 (p(t)))_-$ for the initial state $\ket{\psi (0)_-}$. 

 It is easy to rewrite the bounds (\ref{RQLOSC1}),(\ref{RQLOSC2}) and extremal states (\ref{extremalOSC}) in  dimensionless variables in 
 terms of the dimensional $X$ and $P$ for the oscillator. 
 Thus we have, the RQL for the oscillator,
 \begin{eqnarray}\label{RQLOSCXP}
 &&\cos ^2(\omega t)\sigma^2 (X(0)) +  \frac{\sin ^2(\omega t)}{m^2\omega^2 } \sigma^2 (P(0)) \nonumber \\
&& -\frac{ |\sin (2\omega t)|}{2m\omega } \sqrt{4 \sigma^2(X(0)) \sigma^2(P(0)) -\hbar^2 } \nonumber \\
&& \leq \sigma^2 (X(t))   \nonumber \\
&& \leq \cos ^2(\omega t)\sigma^2 (X(0)) +  \frac{\sin ^2(\omega t)}{m^2\omega^2 } \sigma^2 (P(0)) \nonumber \\
&& +\frac{ |\sin (2\omega t)|}{2m\omega } \sqrt{4 \sigma^2(X(0)) \sigma^2(P(0)) -\hbar^2 }.
\end{eqnarray}

which shows that in the limit $\omega \rightarrow 0$ the RQL for the oscillator ( \ref{RQLOSCXP} ) yields 
the RQL for a free mass ,Eq. (\ref{RQL}).

{\bf Connection of extremal oscillator states with squeezed coherent states}.
The extremal oscillator states have a close connection with squeezed coherent states 
with arbitrary squeezing direction. There are many applications of such optical states in quantum optics \cite{squeezed} and optomechanics.In particular  
there has been progress in preparing a mechanical oscillator in non-Gaussian quantum states \cite{Khalili} by transfering such states from 
optical fields onto the oscillator. Squeezed coherent states have already been utilised in precision measurements needed in 
gravitational interferometers \cite{squeezed measurements}.

  Using  the definitons,
\begin{equation}
  a=\frac{x+ip }{\sqrt{2} }, \> \alpha = < \psi (0)|a \ket{\psi (0)},
\end{equation}
  the  extremal oscillator eigen value equation (\ref{extremalOSC}) 
is equivalent to ,
\begin{eqnarray}
&& (b-\beta)\ket{\psi (0)}=0, \>with \> b = \mu a + \nu a^\dagger, \> \beta= \mu \alpha + \nu \alpha ^*, \nonumber\\
&& \nu/\mu = (\eta-1)/(\eta +1),\> \eta= (\mu  + \nu )/(\mu  - \nu ), \label{TCS1} 
\end{eqnarray}
where we have suppressed the sub-scripts $\pm$ on $\ket{\psi (0)}, \eta, \mu$ and $\nu$ for simplicity. 
Given $\eta$, only the ratio $\nu/\mu $ is fixed; so we can make the convenient choice,

\begin{equation}
 |\mu|^2 - |\nu|^2 =1, \mu >0, \>i.e. \mu = \cosh r ,\> \> \nu =  e^ {i\theta} \sinh r, 
\end{equation}
with $r >0, \theta real $ ,in order to make the transformation from $a,a^\dagger $ to $b,b^\dagger $  canonical, i.e.  $[b,b^\dagger]=1.$.
Eqn. (\ref{TCS1} ) is then just a twisted coherent state  eigen value equation.The unitary displacement operator $D$ and squeeze operator $S$ ,
\begin{eqnarray}
 && D(\beta,b)=D(\alpha,a)=\exp {(\alpha a^\dagger -\alpha^* a) } , \nonumber \\
 && S(\xi)=\exp {\frac{1}{2} \big (\xi^*a^2-\xi a^{\dagger 2} \big) },\> \xi \equiv r \exp {(i\theta ) },
\end{eqnarray}
obey ,
\begin{eqnarray}
&& D^\dagger (\beta,b) b D (\beta,b) = b + \beta ,\nonumber\\
&& S^\dagger(\xi) a S(\xi) =  a \cosh r -a^\dagger  e^ {i\theta} \sinh r \>.
\end{eqnarray}
Defining $a\ket{0}=0 $ we have ,
\begin{eqnarray}
&&(b-\beta) \ket {\alpha,\xi }=0,\> \ket {\alpha,\xi }\equiv D(\alpha,a) S(\xi) \ket {0} , \\
&& i.e.\> \ket{\psi(0) }=\ket {\alpha,r \exp {(i\theta ) } }.
\end{eqnarray}
Thus the extremal states $\ket{\psi(0) } $ are simply related to the squeezed coherent states.
 Since ,
\begin{equation}
 \eta= \frac{1+i \sin{ \theta} \sinh{ (2r)} } { \cosh {(2r)}-\cos {\theta} \sinh { (2r)} },
\end{equation}
$\sin \theta >0$ and $\sin \theta <0$ correspond respectively to $\ket{\psi(0) }_+ $ and $ \ket{\psi(0) }_-$.
 The Heisenberg equations of motion give $ a(t)=a \exp{(-i\omega t ) }$, and hence the time dependent states are,
 \begin{equation}
 \exp {(-iH t) }\ket{\psi(0) }= e^{-i\omega t/2 } \ket {\alpha e^{-i\omega t },r e^ {i(\theta -2\omega t) } }.  
 \end{equation}

{\bf Position Measurements On Free Masses and Harmonic Oscillators Using Contractive States}.
The RQL given above only consider unitary evolution with the system Hamiltonian. Caves \cite{Caves1985} noted insightfully that additional 
considerations involving system-meter interactions during measurement are necessary, and sometimes important.
The von Neumann model \cite{von Neumann} is a prototype of quantum measurement models which couple the system to a meter, 
and monitor the meter position $y$ to obtain information about the system position $x$. Caves  considered a class of models 
(which include the von Neumann model), in which , at any time $\tau$, $\sigma ^2 (y(\tau)) = \sigma ^2 (x(\tau)) +\sigma ^2 _R$, where $ \sigma _R$ 
is the meter resolution. He showed that for measurements  at $t=0$ and $t=\tau$ using identical meter states, the assumption $ \sigma _R \ge \sigma (x(0))$, 
where $\sigma (x(0))$ is the position uncertainty just after the first measurement would again imply the heuristic SQL 
$\sigma^2 (X(\tau)) \geq   \hbar \tau/m $. The SQL also applies to extensions of the Caves \cite{Caves1985} model to 
continuous measurements by  Caves and Milburn, and others \cite{Continuous}.

In order to exploit the new possibilities allowed by the contractive states which violate the SQL (but obey the RQL), 
I outline below the use of the Ozawa interaction Hamiltonian \cite{Ozawa1988} ,
\begin{equation}
 H= k [2x p_y -2p_x y +(x p_x +p_x x -yp_y-p_y y)/2 ],
\end{equation}
where $x,p_x$ are position and momentum operators for the system, and $y,p_y$ those for the meter.
The important properties of this interaction are that ,for a carefully chosen interaction time, after the measurement,
(i) the meter uncertainty  does not contain the additional uncertainty $\sigma_R$ mentioned above ,and (ii) the contractive state of the 
meter is transferred to the system.

Suppose $N$ measurements, each of time duration $\tau$ are made over time intervals 

$$t \epsilon \>[0,\tau], [T,T+\tau],[2T,2T+\tau],...[(N-1)T,(N-1)T +\tau] $$ 

by $N$ meters, each identically prepared at the beginning of the respective measurement in the same contractive 
  state given by Eq.( \ref{chi} )
  \begin{equation}
   <y'\ket {\chi}=\big ( \frac{Re \>\eta _+} {\pi } \big )^{1/4} \exp \big( -\frac{\eta _+ y'^2 } {2}\big ),
  \end{equation}
where we have chosen $<y(0)>=<p_y (0)>=0$ for simplicity, and 
\begin{equation}\label{eta+}
 \eta _{+} = \frac{1}{2 \sigma^2(y(0))}[1 + i \sqrt{4 \sigma^2(y(0))\sigma^2(p_y(0)) -1 }] .
\end{equation}
The meter may for example be an oscillator of frequency $\Omega$ with $\Omega \neq \omega$ where $\omega$ is 
the frequency of the system oscillator.
The coupling strength $k$ is assumed large enough and the time interval $\tau$  small enough for the 
free Hamiltonians of the system and meter to be negligible during these measurement periods.
 
 During each of $N-1$ time intervals of duration $T-\tau$ between successive measurements,
 
 $$ t\epsilon \> [\tau,T],[T+\tau, 2T],...[(N-1)T +\tau, NT]$$,

 the measurement interaction is switched off and the system (free mass or harmonic oscillator) evolves unitarily according to 
its free Hamiltonian. 
At the beginning of each measurement period, (e.g. $t=0,T,2T,..$), i.e. $t=t_i=(i-1)T, i=1,2,..N $, 
the joint wave function of the system and meter is ,
\begin{equation}
 <x',y'\ket {\Psi (t_i)}= <x'\ket {\psi (t_i) } <y'\ket {\chi },
\end{equation}
where we have suppressed a sub-script $i$ referring to the $i-$th meter. Solving the Heisenberg equation of motion using the Ozawa interaction,
we get the operators  after time $\tau$,
\begin{eqnarray}
 x(t_i+\tau)=\frac{2}{\sqrt{3}} \big [\sin {(k\tau\sqrt{3}+\frac{\pi}{3} ) } x(t_i) -\sin {(k\tau\sqrt{3}) } y(t_i) \big ]\nonumber\\
 y(t_i+\tau)= \frac{2}{\sqrt{3}} \big [\sin {(k\tau\sqrt{3}) } x(t_i) +\sin {(\frac{\pi}{3}-k\tau\sqrt{3} ) } y(t_i) \big ],\nonumber
\end{eqnarray}
and the corresponding wave function
 \begin{eqnarray}
 && <x',y'\ket {\Psi (t_i +\tau)}= \nonumber\\
 && <\frac{2}{\sqrt{3}} \big [\sin {(k\tau\sqrt{3}) } y' +\sin {(\frac{\pi}{3}-k\tau\sqrt{3} ) } x'  \big ]\ket {\psi (t_i) } \times\nonumber\\
 && <\frac{2}{\sqrt{3}} \big [\sin {(k\tau\sqrt{3}+\frac{\pi}{3} ) } y' -\sin {(k\tau\sqrt{3}) } x' \big ]\ket {\chi }.
 \end{eqnarray}
If we choose the product of the strength and duration of the interaction such that
\begin{equation}
 k\tau= \pi/ (3\sqrt{3}).
\end{equation}
we get the simple operators and wave functions,
\begin{eqnarray}
&& x(t_i+\tau)= x(t_i) -y(t_i) ; \> y(t_i+\tau)= x(t_i),\\
&& <x',y'\ket {\Psi (t_i +\tau)}= < y'\ket {\psi (t_i) } < y' - x' \ket {\chi }.
\end{eqnarray}
Hence observation of the meter after the measurement will return the correct expectation value for the system before the measurement,  
\begin{equation}
 < \Psi (t_i )|y(t_i +\tau)-x(t_i)\ket {\Psi (t_i )}=0, 
\end{equation}
and the predicted probability density $P(y')$ for the meter,
\begin{equation}
  P(y')(t_i+\tau) = \int dx'|<x',y'\ket {\Psi (t_i +\tau)}|^2=|<y'\ket{\psi(t_i) }|^2 ,
\end{equation}
which is identical to the system position probability density just before measurement. Hence,
\begin{equation}
 \sigma ^2 (y (t_i+\tau) )= \sigma ^2 (x (t_i) ),
\end{equation}
without any extra error $\>\sigma_R\>$ corresponding to meter resolution. Further, after a meter reading $y'$,the system 
 is left in the state 
\begin{equation}
 <x'\ket {\psi (t_i+\tau)}=< y' - x' \ket {\chi } \big[\frac{< y'\ket {\psi (t_i) } }{| < y'\ket {\psi (t_i) } |}\big],
\end{equation}
which, apart from the phase factor in the square bracket on the right-hand side, is just  
the contractive state in which the meter was prepared, but with $<x>=y'$. Using this result and our previous results 
in Eqs. (\ref{tM1},\ref{tM2} ), it follows that the choice
\begin{eqnarray}
&& T-\tau = t_M'\> for\> oscillator; \nonumber\\
&& T-\tau = t_M, \> for \>free \> mass,
\end{eqnarray}
will ensure that the system state has position uncertainty less than the initial meter uncertainty for  $NT > t >\tau$ .
To justify neglecting the free Hamiltonians during the measurement interval $\tau$ we need $  \Omega \tau <<1$ for the meter and 
$ \omega \tau << 1$ if the system is an oscillator, $ \sigma(P(0))/\sigma(X(0)) \tau/m <<1$ if the system is a free mass;
we need the error in the condition $k\tau= \pi/ (3\sqrt{3}) $ to be negligible, i.e.the error $ k \delta \tau <<1$. Hence 
we have the following necessary conditions on the sensitivity of the time setting $\delta \tau$ and the strength $k$ of the 
measurement interaction:
\begin{equation}
 \delta \tau <<\frac{1}{k}=\tau \frac{3\sqrt{3} } {\pi } << min [\frac{1}{\Omega}, \frac{1}{\omega} ],
\end{equation}
for measurements on the oscillator ; for the case of the free mass $1/\omega \rightarrow m \sigma(X(0)/ \sigma(P(0)) $ 
on the right-hand side of the above equation.

{\bf Conclusion}. I have obtained rigorous quantum limits on the variance $\sigma^2 (X(t))$ in terms of  $\sigma^2 (X(0))$ and  $\sigma^2 (P(0))$ 
for arbitrary quantum states of a free mass and of a harmonic oscillator. I also obtained the states which achieve saturation of the limits and 
their connection with squeezed coherent states of an oscillator with arbitrary squeezing direction. In order to utilise the contractive states 
to obtain accuracies beyond the SQL , I have outlined measurement models over finite non-zero time intervals for free mass position and 
oscillator position using the Ozawa Hamiltonian \cite{Ozawa1988} for system-meter interaction.Between measurements the system evolves according to the 
free Hamiltonian. In the oscillator case the extremal contractive state improves on the Schr\"odinger coherent states for a well defined time interval, and 
the free evolution period is adjusted to be equal to that interval.I also briefly discuss the experimental sensitivities needed to justify the 
assumptions on the parameters of the model.

{\bf Acknowledgements}. I thank the Indian National Science Academy for the INSA honorary scientist position at HBCSE, TIFR. I thank 
Priyanshi Bhasin and Ujan Chakraborty who collaborated with me on a related problem \cite {Priyanshi} investigating generalized twisted coherent 
states for a free mass using the generalized oscillator coherent states with non-minimum uncertainty \cite{Roy-Singh}. I thank Paul Busch 
for discussions during the ICQF-17 conference and the suggestion to mention that Eqn.(\ref{Cauchy}) is a rearrangement 
of the Schr\"odinger-Robertson \cite{Kennard} uncertainty relations on position and momentum. I thank N. Mukunda for references to their far 
reaching 'covariant' generalizations of the  Schr\"odinger-Robertson uncertainty relations to systems with any number 
of degrees of freedom.\cite{Mukunda}.
I also thank Alok Pan and Archan Majumdar for inviting me to present these results respectively at 
the ICQF-17 conference at NIT, Patna and the ISNFQC18 conference at SN Bose Centre, Kolkata .
I thank the referee for valuable suggestions on connections with squeezed coherent states and 
on the importance of non-instantaneous measurement models.

\end{document}